\documentclass[12pt]{article}
\pdfoutput=1
\usepackage{subfigure}
\usepackage{amssymb,amsmath}
\usepackage{graphicx}
\usepackage{color}
\usepackage[colorlinks=true
,urlcolor=blue
,citecolor=blue
,linkcolor=blue
,pagecolor=blue
,linktocpage=true
,pdfproducer=medialab
]{hyperref}
\usepackage[a4paper,width=15.33cm]{geometry}
\makeatletter \renewcommand{\@dotsep}{10000} \makeatother

\newcommand{\beq}{\begin{equation}}
\newcommand{\eeq}{\end{equation}}
\newcommand{\bea}{\begin{eqnarray}}
\newcommand{\eea}{\end{eqnarray}}

\begin{document}
%Remove date before submitting to arXiv
%\date{\today}

%\baselineskip 36pt
\begin{center}

 {\large\bf    Higgs and Sparticle Masses from Yukawa Unified SO(10):\\ A Snowmass White Paper
  } \vspace{1cm}

{\large   M. Adeel Ajaib\footnote{ E-mail: adeel@udel.edu}, Ilia Gogoladze\footnote{E-mail: ilia@bartol.udel.edu\\
\hspace*{0.5cm} On  leave of absence from: Andronikashvili Institute
of Physics,  Tbilisi, Georgia.},  Qaisar Shafi\footnote{ E-mail:
shafi@bartol.udel.edu} and Cem Salih $\ddot{\rm U}$n \footnote{
E-mail: cemsalihun@bartol.udel.edu}} \vspace{.5cm}

{\baselineskip 20pt \it
Bartol Research Institute, Department of Physics and Astronomy, \\
University of Delaware, Newark, DE 19716, USA  } \vspace{.5cm}

%\vspace{1.5cm}
\end{center}

\begin{abstract}

We discuss ways to probe $t-b-\tau$ Yukawa coupling unification condition at the Energy and Intensity frontiers. We consider non-universal soft supersymmetry breaking mass {terms} for gauginos related by the SO(10) grand unified theory (GUT). We have previously shown that $t-b-\tau$ Yukawa coupling unification prefers a mass of around 125 GeV for the Standard Model-like Higgs boson with all colored sparticle masses above 3 TeV. The well-known MSSM parameter $\tan\beta \approx 47-48$ and neutralino-stau coannihilation yields the desired relic dark matter density.

\end{abstract}

\vspace{.8cm}

%\newpage

%%%%%%%%%%%%%%%%%%%%%%%%%%%%%%%%%%%%%%%%%%%%%%%%%%%%%%%%%%%%
\renewcommand{\thefootnote}{\arabic{footnote}}
\setcounter{footnote}{0}

%%%%%%%%%%%%%%%%%%%%%%%%%%%%%%%%%%%%%%%%%%%%%%%%%%%%%%%%%%%%%

%\baselineskip 36pt
% Main body
%%%%%%%%%%%%%%%%%%%%%%%%%%
%\baselineskip 18pt
%%%%%%%%%%%%%%%%%%%%%%%%%%

The discovery of a Standard Model (SM)-like Higgs boson with mass $\sim$ 125 GeV by the ATLAS \cite{:2012gk} and the CMS \cite{:2012gu} collaborations
 is compatible with low ($\sim 1-$few TeV) scale supersymmetry.
 In addition to solving the gauge hierarchy problem and providing a compelling dark matter candidate, supersymmetry also leads to unification of the three SM gauge couplings at the grand unification scale ($M_{{\rm GUT}} \sim 10^{16} $ GeV). Furthermore $t-b-\tau$  Yukawa \cite{big-422} coupling unification at $M_{{\rm GUT}}$ arises in supersymmetric SO(10), in  contrast to its non-supersymmetric counterpart. In Ref. \cite{Gogoladze:2011aa} we considered a supersymmetric SO(10) model with  non-universal  soft supersymmetry breaking (SSB) gaugino masses at $M_{{\rm GUT}}$, arising from a non-singlet
F-component of the field which breaks supersymmetry. We discussed, in particular, a gravity mediated supersymmetry breaking scenario \cite{Chamseddine:1982jx}  with  the following relation at $M_{{\rm GUT}}$
among the masses of the minimal supersymmetric standard model (MSSM),
\begin{align}
M_3: M_2:M_1= -2:3:1 ,
\label{gaugino10}
\end{align}
where $M_1, M_2, M_3$ denote the gaugino masses of $U(1)$, $SU(2)_L$ and $SU(3)_c$
respectively.   In order to obtain the correct sign for the desired contribution to $ (g-2)_{\mu} $, we set $\mu > 0$, $ M_{1} > 0 $, $ M_{2} > 0 $ and $ M_{3} < 0 $.

Employing the boundary condition from Eq.(\ref{gaugino10}), one  can define the MSSM gaugino masses at $ M_{\rm GUT} $ in terms of the mass parameter $M_{1/2}$:
\begin{align}
M_1= M_{1/2},~~~~~~~
M_2= 3M_{1/2}~~~~~~~ \rm{and}~~~~~~~
M_3= - 2 M_{1/2}.
 \label{gaugino11}
\end{align}
In order to quantify Yukawa coupling unification, we define the quantity $R_{tb\tau}$ as,
\begin{align}
R_{tb\tau}=\frac{ {\rm max}(y_t,y_b,y_{\tau})} { {\rm min} (y_t,y_b,y_{\tau})}.
\end{align}

We have performed random scans for the following parameter range:
\begin{align}
0\leq  m_{16}  \leq 10\, \rm{TeV} \nonumber \\
0\leq   m_{10} \leq 10\, \rm{TeV} \nonumber \\
0 \leq M_{1/2}  \leq 5 \, \rm{TeV} \nonumber \\
35\leq \tan\beta \leq 55 \nonumber \\
-3\leq A_{0}/m_{16} \leq 3.
 \label{parameterRange}
\end{align}
 Here $ m_{16} $ is the universal SSB mass for MSSM sfermions, $ m_{10} $ is the universal SSB mass term for up and down MSSM Higgs masses, $ M_{1/2} $ is the SSB gaugino mass parameter, $ \tan\beta $ is the ratio of the vacuum expectation values (VEVs) of the two MSSM Higgs doublets, $ A_{0} $ is the universal SSB trilinear scalar interaction (with corresponding Yukawa coupling factored out).  We use the central value $m_t = 173.1\, {\rm GeV}$  \cite{:1900yx} and with 1$\sigma$ deviation in our analysis.
We use $m_b(m_Z)=2.83$ GeV which is hard-coded into Isajet.

We employ Isajet~7.84 \cite{ISAJET}  interfaced with Micromegas 2.4 \cite{Belanger:2008sj} to perform random
scans over the fundamental parameter space.
Micromegas is interfaced with Isajet  to calculate the relic density and branching ratios $BR(B_s \rightarrow \mu^+ \mu^-)$ and $BR(b \rightarrow s \gamma)$. With these codes we implement the following random scanning procedure: A uniform and logarithmic distribution of random points is first generated in the parameter space given in Eq.(\ref{parameterRange}). A Gaussian distribution of points is then generated around each point in the parameter space.
  We successively apply the following experimental constraints on the data that
we acquire from this interface:
\begin{table}[h!]\centering
\begin{tabular}{rlc}
$ 0.8 \times 10^{-9} \leq BR(B_s \rightarrow \mu^+ \mu^-) $&$ \leq\, 6.2 \times 10^{-9} \;
 (2\sigma)$        &   \cite{:2007kv}      \\
$2.99 \times 10^{-4} \leq BR(b \rightarrow s \gamma) $&$ \leq\, 3.87 \times 10^{-4} \;
 (2\sigma)$ &   \cite{Barberio:2008fa}  \\
$0.15 \leq \frac{BR(B_u\rightarrow
\tau \nu_{\tau})_{\rm MSSM}}{BR(B_u\rightarrow \tau \nu_{\tau})_{\rm SM}}$&$ \leq\, 2.41 \;
(3\sigma)$ &   \cite{Barberio:2008fa}  \\
 $ 0 \leq \Delta(g-2)_{\mu}/2 $ & $ \leq 55.6 \times 10^{-10} $ & \cite{Bennett:2006fi}
\end{tabular}\label{table}
\end{table}

%%%%%%%%%%%%%%%%%%%%%%%%%%%%%%%%%%%%%%%%%%%%%%%%%%%%%%%%%%%
%trim=l b r t
%\newpage
\begin{figure}[]
\newpage
%\vspace{-1cm}
\centering
\subfigure{\label{fig:3-a}{\includegraphics[scale=0.4]{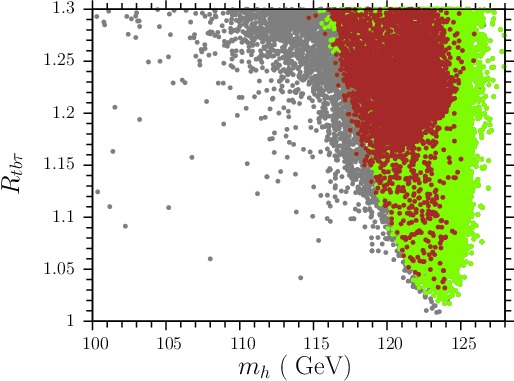}}}
\caption{Plot in the $R_{tb\tau} - m_h$ plane.
Gray points are consistent with REWSB and neutralino LSP. Green points satisfy particle mass bounds and constraints from $BR(B_s\rightarrow \mu^+ \mu^-)$, $BR(b\rightarrow s \gamma)$
and $BR(B_u\rightarrow \tau \nu_\tau)$. In addition, we require that green points do no worse than the SM in terms of $(g-2)_\mu$. Brown  points belong to a subset of green points and satisfy the WMAP bounds ($\Omega h^2<1 $) on neutralino dark matter abundance.}
\label{fig:R-mh}
\end{figure}

%%%%%%%%%%%%%%%%%%%%%%%%%%%%%%%%%%%%%%%%%%%%%%\vspace*{2mm}

%trim=l b r t
%\newpage
\begin{figure}[]
\newpage
%\vspace{-1cm}
\centering
{\label{fig:isa-d}{\includegraphics[scale=0.4]{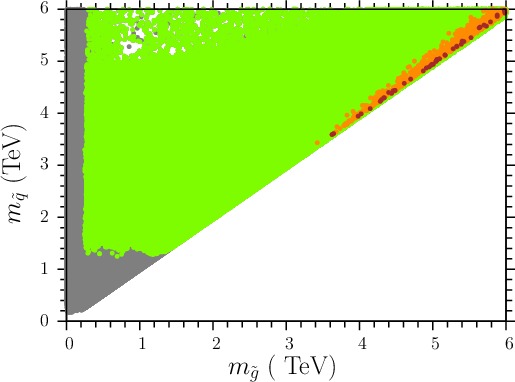}}}\hfill
{\label{fig:isa-e}{\includegraphics[scale=0.4]{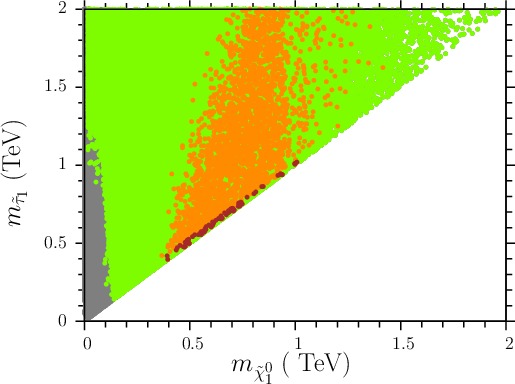}}}
\caption{Plots in $m_{\tilde{q}}-m_{\tilde{g}}$, $m_{\tilde{\tau}_1}-m_{\tilde{\chi}^0}$ planes. The color coding is the same as that used in Figure \ref{fig:R-mh}. In addition, yellow points correspond  to Yukawa unification better than $ 10\% $. Brown points  are compatible with WMAP bound on relic abundance ($\Omega h^2<1 $) are a subset of yellow points.}
\label{fig2}
\end{figure}

In Figure  \ref{fig:R-mh}, we present results in the $ R_{tb\tau}-m_{h} $ plane. Gray points are consistent with radiative electroweak symmetry breaking (REWSB) and neutralino being lightest supersymmetric particle (LSP).
The green points satisfy particle mass bounds and constraints from $ BR(B_{s}\rightarrow \mu^{+}\mu^{-}) $, $ BR(b\rightarrow s \gamma) $ and $ BR(B_{u}\rightarrow \tau\nu_{\tau}) $. In addition, the green points do no worse than the SM in terms of $ (g-2)_{\mu} $. The brown points belong to a subset of the green points and satisfy the WMAP bound ($\Omega h^2 < 1$) on neutralino dark matter abundance.  It is exciting to note that demanding exact $t-b-\tau$ Yukawa unification in the given supersymmetric SO(10) GUT allows us to predict a mass ($m_h$) of around 124 GeV for the SM-like Higgs boson, consistent with all collider and astrophysical constraints  mentioned above.
 This prediction is in very good agreement with the recent ATLAS and CMS measurements \cite{:2012gk,:2012gu}.
 Requiring $t-b-\tau$ Yukawa unification at $5\%$ level we find the result $122\,{\rm GeV} < m_h < 126\,{\rm GeV}$.
 
Note that there is approximately a $\pm$2 GeV theoretical uncertainty in the calculation of this mass. It is interesting to note that the Higgs mass interval is significantly reduced compared with versions of the SO(10) and 4-2-2 models with $m_{H_{u}}\neq m_{H_{d}}$ for which the SM-like Higgs boson mass  lies in the interval $80~ {\rm GeV} < m_h < 129$ GeV \cite{Gogoladze:2009ug}, compatible with exact $t-b-\tau$ Yukawa unification. 

Note that  in a class of {models,} where {a non-trivial} asymptotic relation  among the third generation quark and lepton Yukawa couplings can be derived
\cite{Ajaib:2013kka,Gomez:2002tj}, we still have a {relatively} big mass interval for the SM-like Higgs mass.

%%%%%%%%%%%%%%%%%%%%%%%%%%%%%%%%%%%%%%%%%%%%%%\vspace*{2mm}
\begin{table}[p!]
\centering
%\begin{tabular}{|p{3cm}|p{3cm}|p{3cm}|p{3cm}|p{3cm}|}
\begin{tabular}{|c|c|c|c|c|}
\hline
\hline
                 	&	 Point 1  	&	 	 Point 2  	&	 Point 3 	\\
%\times 10^{									
\hline									
%\times 10^{- \times 10^{ \times 10^{									
									
$m_{10} $         	&$	          4.19\times 10^{2}	$&$	         	          4.49\times 10^{2}	$&$	          1.94\times 10^{3}	$\\
$m_{16} $         	&$	          2.13\times 10^{3}	$&$	         	          1.91\times 10^{3}	$&$	          2.00\times 10^{3}	$\\
$M_1$         	&$	          1.89\times 10^{3}	$&$	                   1.78\times 10^{3}	$&$	          1.51\times 10^{3}	$\\
$M_2$         	&$	          5.67\times 10^{3}	$&$	          	          5.35\times 10^{3}	$&$	          4.53\times 10^{3}	$\\
$M_3$         	&$	         -3.78\times 10^{3}	$&$	         	         -3.57\times 10^{3}	$&$	         -3.02\times 10^{3}	$\\
$A_0/m_{16}$         	&$	2.39		$&$	0.03	$&$	1.56	$\\
$\tan\beta$      	&$	47.18		$&$	47.93	$&$	47.46	$\\
$m_t$            	&$	174.2		$&$	174.2	$&$	173.1	$\\
\hline		  		  		  		  	
$\mu$            	&$	3729		$&$	2913	$&$	2526	$\\

\hline		  		  		  		  	
$m_h$            	&$	125		$&$	124	$&$	123	$\\
$m_H$            	&$	747		$&$	572	$&$	558	$\\
$m_A$            	&$	742		$&$	568	$&$	554	$\\
$m_{H^{\pm}}$    	&$	753		$&$	580	$&$	567	$\\
		  		  		  		  	
\hline		  		  		  		  	
$m_{\tilde{\chi}^0_{1,2}}$	&$	         895,         3739	$&$	         	         848,         2932	$&$	         709,         2540	$\\

$m_{\tilde{\chi}^0_{3,4}}$	&$	        3742,         4822	$&$	        	        2935,         4562	$&$	        2543,         3849	$\\

$m_{\tilde{\chi}^{\pm}_{1,2}}$	&$	        3789,         4774	$&$	        	        2978,         4516	$&$	        2579,         3809	 $\\

$m_{\tilde{g}}$  	&$	7694		$&$	7266	$&$	6239	$\\
		  		  		  		  	
\hline $m_{ \tilde{u}_{L,R}}$	&$	        7667,         6824		$&$	        7219,         6415	$&$	        6295,         5635	 $\\
                 		  		  		  		  	
$m_{\tilde{t}_{1,2}}$	&$	        5331,         6560	$&$	        	        5239,         6367	 $&$	        4390,         5370	$\\
                 		  		  		  		  	
\hline $m_{ \tilde{d}_{L,R}}$	&$	        7668,         6814	$&$	        	        7220,         6406	$&$	        6296,         5628	 $\\
                 		  		  		  		  	
$m_{\tilde{b}_{1,2}}$	&$	        5553,         6526	$&$	        	        5434,         6333	 $&$	        4591,         5341	$\\
                 		  		  		  		  	
\hline		  		  		  		  	
$m_{\tilde{\nu}_{1,2}}$	&$	4148		$&$	3870	$&$	3487	$\\
                 		  		  		  		  	
$m_{\tilde{\nu}_{3}}$	&$	3898		$&$	3641	$&$	3243	$\\
                 		  		  		  		  	
\hline		  		  		  		  	
$m_{ \tilde{e}_{L,R}}$	&$	        4153,         2234		$&$	        3875,         2009	 $&$	        3491,         2068	$\\
                		  		  		  		  	
$m_{\tilde{\tau}_{1,2}}$	&$	           1094,   3875   		$&$	  881,      3620 $&$	    1061,    3225     	 $\\
                		  		  		  		  	
\hline		  		  		  		  	
$\Delta(g-2)_{\mu}$  	&$	  3.11\times 10^{-11}	$&$	 	  3.71\times 10^{-11}	$&$	  4.97\times 10^{-11}	$\\

$\sigma_{SI}({\rm pb})$	&$	  1.59\times 10^{-11}	$&$	  	  7.08\times 10^{-11}	$&$	  1.00\times 10^{-10}	$\\

$\sigma_{SD}({\rm pb})$	&$	  4.69\times 10^{-10}	$&$	  1	  1.60\times 10^{-9}	$&$	  2.89\times 10^{-9}	$\\

$\Omega_{CDM}h^{2}$	&$	  6.5	$&$	  	  0.8	$&$	  4.0	$\\
                		  		  		  		  	
\hline		  		  		  		  	
		  		  		  		  	
$R_{t b \tau}$     	&$	1.02		$&$	1.03	$&$	1.04	$\\

\hline
\hline
\end{tabular}
\caption{Benchmark points with good Yukawa unification. All the masses are in units of GeV. Point 1  demonstrates how a small value of $R_{t b \tau}$ yields a Higgs mass $\sim 125$ GeV. Point 2 exhibits stau coannihilation and has a small $R_{t b \tau}$ that agrees with $\Omega h^2 < 1$. Point 4 shows that good $t-b-\tau$ Yukawa unification can be attained with the sfermions and Higgs nearly degenerate ($m_{16} \simeq m_{10}$) at $M_{GUT}$. }
\label{tab1}
\end{table}

In Figure \ref{fig2}  the color coding is the same as  in Figure \ref{fig:R-mh}. In addition, yellow points correspond  to Yukawa unification better than $ 10\% $. Brown points  are compatible with the WMAP bound on relic abundance ($\Omega h^2<1 $) are a subset of yellow points.
The $m_{\tilde{q}}-m_{\tilde{g}}$ panel shows that $t$-$b$-$\tau$ Yukawa unification in our scenario predicts masses for the gluino and the first two family squarks which lie  around 4 TeV or so, which can be observed in the future at LHC.

In the present framework, the WMAP constraint on the relic dark matter abundance is only satisfied by the neutralino-stau coannihilation scenario where the neutralino and stau masses are degenerate to a good approximation (see brown points).
 From the plot in $ m_{\tilde{\tau}}-m_{\tilde{\chi}^{0}_1} $ plane of Figure \ref{fig2}, we observe that demanding $10\%$  or better Yukawa unification yields the following constraint on the LSP neutralino mass, $400\, {\rm GeV} \lesssim  m_{\tilde{\chi}^{0}_1} \lesssim 1\, {\rm TeV}$. It remains to be seen whether alternative coannihilation scenarios may emerge after a more through analysis.

In Table \ref{tab1} we present three benchmark points with acceptable degree of Yukawa unification and SM-like Higgs mass close to 125 GeV (recall the $\pm$ 2 GeV theoretical uncertainty in estimating this mass). The points shown also satisfy the constraints described above. {Point} 1  represent solutions that yield almost perfect Yukawa unification (of around $2\%$).  Point 2 depicts stau coannihilation in addition to a 124 GeV Higgs and $R_{tb\tau}=1.03$. Point 3 shows that perfectly respectable $t-b-\tau$ Yukawa unification can be achieved with nearly degenerate ($m_{16} \simeq m_{10}$) sfermion and Higgs scalar masses at $M_{GUT}$.

%%%%%%%%%%%%%%%%%%%%%%%%%%%%%%%%%%%%%%%%%%%%%%
\section*{Acknowledgments}

This work is supported in part by the DOE Grant No. DE-FG02-91ER40626. This work used the Extreme Science
and Engineering Discovery Environment (XSEDE), which is supported by the National Science
Foundation grant number OCI-1053575.

%%%%%%%%%%%%%%%%%%%%%%%%%%%%%%%%%

\end{document}